\documentclass[dvips,12pt,a4paper]{article}
\usepackage{amsmath,pifont,amsfonts,amssymb,latexsym,multicol}
\usepackage{float,xspace,calc,theorem,ifthen}
\usepackage{fancyhdr,enumerate,psboxit,epsfig}
\usepackage{times,amscd,epic,epsfig}
\usepackage{version,xr}
\usepackage{hyperref}

\newtheorem{theorem}{Theorem} [section]
\newtheorem {lemma}[theorem]{Lemma}
\newtheorem {prop}[theorem]{Proposition}
\newtheorem {definition}[theorem]{Definition}

{\theorembodyfont{\normalfont}\newtheorem {remark}[theorem]{Remark}}
{\theorembodyfont{\normalfont}\newtheorem {remarks}[theorem]{Remarks}}
\numberwithin{equation}{section}
\newcommand{\beq}{\begin{equation}}
\newcommand{\eeq}{\end{equation}}
\newcommand{\Leq}[1]{\label{#1}\end{equation}}
\newcommand{\beqn}{\begin{eqnarray}}
\newcommand{\eeqn}{\end{eqnarray}}
\newcommand{\beqno}{\begin{eqnarray*}}
\newcommand{\eeqno}{\end{eqnarray*}}
\renewcommand {\l}{\left}
\newcommand {\ri}{\right}
\newcommand {\s}{{s}}
\newcommand {\q}{{q}}
\newcommand {\p}{{p}}
\newcommand {\Q}{{Q}}

\newcommand {\pq}{{(\p,\q)}}

\newcommand {\Rvir}{R_{\rm vir}}       
\newcommand {\rmin}{{r_{\rm min}}}     

\newcommand {\Vmax}{V_{\rm max}}       

\newcommand {\Dir}{{\hat{q}}}
\newcommand {\Dist}{{\rm Dist}}

\newcommand {\hp}{\hat{p}}
\newcommand {\hq}{\hat{q}}

\newcommand {\vep}{\varepsilon}

\newcommand {\vv}{\varphi}

\newcommand {\LA}{\left\langle}
\newcommand {\RA}{\right\rangle}
\newcommand {\pa}{\partial}

\newcommand {\eh}{{\textstyle \frac{1}{2}}}

\newcommand {\ar}{\rightarrow}
\newcommand {\sign}{{\rm sign}}

\newcommand {\SuE}{\Sigma_{E}}
\newcommand {\SuEu}{\SuE^{\rm u}}

\newcommand {\bR}{{\mathbb R}}
\newcommand {\bN}{{\mathbb N}}
\newcommand {\bZ}{{\mathbb Z}}

\newcommand{\rstr}{{\upharpoonright}}
\newcommand{\idty}{{\rm 1\mskip-4mu l}} 

\newcommand{\cI}{{\cal I}}
\newcommand{\cT}{{\cal T}}

\newcommand{\bem}{\l(\! \begin{array}}
\newcommand{\eem}{\end{array}\!\ri)}
\newcommand{\NN}{\nonumber}
\newcommand {\qmbox}[1]{\quad\mbox{#1}\quad}
\newcommand {\IZ}{{\cal I\!Z}} 
\newcommand {\NT}{{\cal N\!T}} 

\newcommand{\cN}{{\cal N}}

\newcommand {\cR}{{\cal R}} 

\begin{document}
\title{The Non-Trapping Degree of Scattering}
\author{Andreas Knauf\and
   Markus Krapf 
   \thanks{Mathematisches Institut der
   Universit\"at Erlangen-N\"urnberg. Bismarckstr. 1 1/2,
   D-91054 Erlangen, Germany.
   e-mail: knauf@mi.uni-erlangen.de }}
\date{\today}

\maketitle
\begin{abstract}
We consider classical potential scattering. If at
energy $E$ no orbit is trapped, 
the Hamiltonian dynamics defines an integer-valued topological degree
${\rm deg}(E)\le 1$. This is calculated explicitly for all
potentials, and exactly the 
integers $\le 1$ are shown to occur for suitable potentials.
 
The non-trapping condition is  
restrictive in the sense that for a bounded potential 
it is shown to imply that the boundary of Hill's Region 
in configuration space is either 
empty or homeomorphic to a sphere.

However, in many situations one can decompose a potential into
a sum of non-trapping potentials with non-trivial degree
and embed symbolic dynamics of multi-obstacle scattering.
This comprises a large number of earlier results, obtained
by different authors on multi-obstacle
scattering.

\end{abstract}
%
\section{Introduction}
%
In potential scattering on $\bR^d$ one considers the solutions
of the Hamiltonian equations for the Hamiltonian function
$H\pq=\eh\|\p\|^{\,2}+V(\q)$, where the {\em potential} $V\in C^2(\bR^d,\bR)$ 
decays at spatial infinity (see (\ref{finite:acc}) below), for positive values $E$ of $H$.
Equivalently one considers the solutions of Newton's equation
$\ddot{q}=-\nabla V(q)$.
The orbit through $x$ is called {\em scattering} if 
$\lim_{|t|\ar\infty} \|q(t,x)\|=\infty$.

Here we consider {\em energies} $E$ for which there are no {\em trapped} 
orbits, i.e.\ solutions where the above limit exists only in one time direction.
Then asymptotically the solutions have the form of straight lines and can
thus be parametrized by a point in the cotangent bundle $N:=T^*S^{d-1}$.
Dynamics induces a diffeomorphism 
\[S_E:N\ar N.\]
In \cite{Kn} this {\em scattering map} was used to define
a topological index, ${\rm deg}(E)\in\bZ$. In examples of centrally symmetric
$V$ all values $\le1$  were shown to occur. 

In Sect.\ \ref{sect:LRP} we begin by defining the
class of potentials for which we can explicitly calculate the index
in the non-trapping case.
This comprises nearly all potentials,
see Thm.\ \ref{thm:reg} for bounded potentials 
and Thm.\ \ref{thm:sing} for potentials with singularities,
like the Coulomb potential. Billiards can be treated by the same
method, see Remarks \ref{rem:billard}

The index is related to the way a Lagrange manifold 
folds over configuration space (Thm.\ \ref{thm:deg:2}).
This shows that only the values $\le1$ can occur.

Finally, in Sect.\ \ref{sect:cobordism} we find in all dimensions $d$ topological obstructions
for $(V,E)$ to lead to motion without trapping.

In \cite{Kn} this index was used to imbed symbolic 
dynamics for scattering in a potential $V=V_1+\ldots+V_k$
for energy $E$, where the
$V_i$ were only assumed to carry non-zero degree,
and to have {\em non-shadowing} supports (no line meeting more than two
supports).
More precisely, for any bi-infinite sequence $a$ in
\beq
\big\{a\in \{1,\ldots,k\}^\bZ\mid a_l\neq a_{l+1}\big\},
\Leq{def:shift}
there exists an orbit of energy $E$, 
visiting the supports of the $V_i$ in the succession 
prescribed by $a$. So the flow has positive topological entropy if
$k>2$.

With the present work, we need not assume any more that the 
building blocks $V_k$ are centrally symmetric, in order to calculate 
their degree and to combine them as indicated above. 

The phenomenon of trapping by chaotic repellers
has been observed and analyzed in many cases, see e.g.\
Rapoport and Rom-Kedar \cite{RR} and references cited therein.
The present work, together with \cite{Kn}, 
provides a unifying approach to several of these results.

Trapping plays a major role in semiclassical quantum
mechanics, and leads to the phenomenon of resonances.
See e.g.\ Castella, Jecko and  Knauf \cite{CJK}, and references cited therein.\\[2mm]
{\bf Acknowledgement:}
We thank Christoph Schumacher (Erlangen) and the anonymous referees 
for helpful comments.

%
\section{Scattering for Long Range Potentials}\label{sect:LRP}
%
We start by introducing the notions of potential scattering and defining 
the topological degree. 

The {\em configuration space} of the scatterer is $\bR^d$, but due
to singularities the
domain of definition $M$ of the potential $V$ may be smaller.\\
For physical and mathematical reasons
we consider potentials $V\in C^2(M,\bR)$
for $M:=\bR^d$ and $M:=\bR^d\setminus\{\s\}$
for some $\s\in\bR^d$ (in Remarks \ref{rem:tra} we also 
consider the case of several singularities).\\ 
In the '{\em singular}' case $M =\bR^d\setminus\{\s\}$ 
we assume that for some $Z>0$, 
$\alpha>0$ and $W\in C^2(\bR^d,\bR)$
\beq
V(\q)=  \frac{-Z}{\|\q-\s\|^\alpha} + W(\q).
\Leq{sing:V} 
Similar to Derezi\'{n}ski and G\'{e}rard \cite{DG}, Sect.\ 2.7 the 
{\em force field} $F:=-\nabla V$ of $V$ is assumed to
meet the {\em long range} estimates for multi-indices $m\in\bN_0^d$ 
\beq
\int_R^\infty \sup_{\|\q\|\ge r} \|\pa^m F(\q)\| \, r^{|m|}\,dr < \infty
\qquad (|m|\le 1)
\Leq{finite:acc}
for some $R$ (say $R= 0$ in the non-singular, $R=2\|s\|$ in the singular case).
\begin{remarks}
\begin{enumerate}
\item
For $d\ge2$ the
long range condition (\ref{finite:acc}) 
implies the existence of $\lim_{\|\q\|\ar\infty}V(\q)$, 
which we assume to be zero without loss of generality.

Evaluated for $m=0$, condition (\ref{finite:acc}) 
leads to finite total change of velocity of the scattered particle
(see Thm.\ 2.5.2 of \cite{DG}). 
If one would want to define so-called M\o ller transformations, 
comparing the dynamics with the one for $V=0$, a {\em short range} 
condition 
\beq
\int_R^\infty \sup_{\|\q\|\ge r} \|\pa^m F(\q)\| \, r^{|m|+1}\,dr < 
\infty \qquad (|m|\le 1)
\Leq{s:r}
would be needed (see Thm.\ 2.6.1 of \cite{DG}).\\ 
But here we neglect time parametrization of the orbits
and consider scattering on a reduced phase space $N$. 
\item
For all values 
$\alpha>0$ the potential $q\mapsto \frac{-Z}{\|\q-\s\|^\alpha}$
meets the long range condition~(\ref{finite:acc}), and for
$\alpha>1$ the short range condition~(\ref{s:r}).

But for all values 
$\alpha\ge2$ under the influence of this potential
the set of initial conditions 
leading to a collision with the singularity in finite
time has positive Liouville measure, see, e.g.\ \cite{LL}, \S 18. We thus assume $\alpha\in (0,2)$.
\end{enumerate}
\end{remarks}
We now consider the Hamiltonian function
\beq
H\in C^2(T^* M,\bR)\qmbox{,}H\pq:=\eh\|\p\|^{\,2}+V(\q)
\Leq{def:H}
on the symplectic manifold $(T^*M,\omega_0)$, with canonical 
symplectic form $\omega_0:=\sum_{k=1}^d dq_k\wedge dp_k$.
\begin{itemize}
\item
For the non-singular case the Hamiltonian flow generated by $H$ 
on the phase space $P:=T^* M$ is complete (see, e.g.\ Sect.\ 2.2 of
\cite{DG}).
\item
Likewise in the singular case it is known that 
precisely for $\alpha=2n/(n+1)$, $n\in\bN$ 
the motion can be regularized. 
For the case of vanishing additional potential term $W$  
in (\ref{sing:V}) this was treated by McGehee in \cite{MG}.

For the general case we obtain in Prop.\ \ref{deg:sing} below 
a complete flow on a
$2d$-dimensional symplectic manifold $P$  which (as a set) equals
\[P = T^* M \,\dot{\cup} \,(\bR\times S^{d-1}).\] 

Physically most important is the case of Coulomb potentials ($n=1$).
\end{itemize}
In both cases we obtain a flow $\Phi\in C^1(\bR\times P,P)$, also denoted by 
\[\Phi^t:P\ar P\qmbox{or}
\big(\p(t,x),\q(t,x)\big) := \Phi^t(x) \qquad(t\in\bR),\]
restricting to the {\em energy shells} $\SuE:=H^{-1}(E)$. 
Moreover  
\[\Vmax:=\sup_{q\in M} V(\q)\ \in[0,\infty).\]

\begin{remark}
In potential scattering, the {\em virial identity}
\beq
\eh\frac{d^2}{dt^2}\| q(t)\|^2 =\frac{d}{dt} \LA\q(t),\p(t) \RA = 2(E-V(\q(t)))- \LA\q(t),\nabla V(\q(t))\RA 
\Leq{virial}
holds true for any trajectory 
$t\mapsto \big(\p(t),\q(t)\big)\equiv\big(\p(t,x),\q(t,x)\big)$, 
with energy $E := H(x)$ (whenever $\q(t)\in M$).
For $E>0$ as a consequence of  (\ref{finite:acc}), 
there exists a {\em virial radius} $\Rvir\equiv\Rvir(E)\ge R$, with
\beq
|V(\q)|< {E}/{2} \qmbox{and} |\LA\q, \nabla V(\q)\RA| < E/2
\qquad(\|\q\|\geq \Rvir). 
\Leq{V:small}
Then by (\ref{virial}) and (\ref{V:small})
\beq
\frac{d}{dt} \LA\q(t),\p(t) \RA > \frac{E}{2} > 0 \qquad\mbox{if }
 \|q(t)\|\geq \Rvir.
\Leq{qp}
Thus a configuration space 
trajectory $t\mapsto\q(t)$ of energy $E$ 
leaving the ball $\IZ(E)\subset\bR^d$ of radius $\Rvir(E)$ (the {\em interaction zone})  
cannot reenter $\IZ(E)$ in the future but goes to spatial infinity.
Namely assume that $\LA\q(0),\p(0)\RA\geq 0$. By (\ref{qp})
\[\frac{d^2}{dt^2}\|\q(t)\|^{2} = 
2\frac{d}{dt}\LA\q(t),\p(t)\RA > E\qquad (t\geq 0)\]
so that 
\beq
\|\q(t)\|^{2} \geq \|\q_0\|^{2} + \eh Et^2 \quad (t\geq 0).
\Leq{d2qge}

Thus after having shown existence of a flow $\Phi\in C^1(\bR\times P,P)$
we can use results, derived in \cite{DG} for scattering by
non-singular potentials, in the singular case, too.
In particular, we have 
\[\limsup_{t\ar+\infty} \|\q(t,x)\| = \infty
\qmbox{if and only if}\lim_{t\ar+\infty} \|\q(t,x)\|= \infty,\]
and similarly for $t\ar-\infty$.
\end{remark}

For $E>0$ {\em Hill's region}
\[\cR_E :=\{ \q \in  M \mid V(\q)\leq E\}\]
is non-empty, but need not be connected 
(since there may be potential pits).

By the assumption $\lim_{\|q\|\ar\infty} V(q)=0$, $\cR_E$ contains
the neighbourhoods of infinity of the form $\{q\in M\mid \|q\|>R\}$
for $R>0$ large.
These are connected if and only if $d\ge2$.
So for $d\geq2$
there is precisely one unbounded connected component $\cR_E^u$ 
of $\cR_E$, and the same is true
for the energy shell $\SuE$ projecting to Hill's region. We denote the
unbounded connected component of $\SuE$ by $\SuEu$.

\begin{definition}\label{def:nt}
$\bullet$
We call $E>0$ a {\bf non-trapping energy} if no orbit in $\SuE$
is {\bf trapped}, that is, for no initial condition $x\in\SuE$
\beq
\lim_{t\ar-\infty} \|\q(t,x)\| = \infty\qmbox{but}
\limsup_{t\ar+\infty} \|\q(t,x)\| < \infty.
\Leq{trap}
$\bullet$
The set of non-trapping energies $E\in(0,\infty)$
is denoted by $\NT$. 
\end{definition}
\begin{remarks}\label{rem:tra}
\begin{enumerate}
\item
Unlike in Def.\ \ref{def:nt}, 
in Def.\ 2.1.3 of \cite{DG}, $E$ is called {\em trapping} if there
exist orbits in $\SuE$ bounded {\em at least}
in the future. \\
That definition has some advantages in the context of 
semiclassical quantum mechanics, but it would unnecessarily
narrow the scope
of our results.
\item
As $\Phi$ is reversible, Def.\ \ref{def:nt} does not change
under a sign change in (\ref{trap}).
\item
Trivially trapped and scattering orbits only occur in $\SuEu$
whereas the orbits in $\SuE\setminus \SuEu$ are bounded, but 
there may be bounded orbits in $\SuEu$ as well.
\item
As shown in Prop.\ 1 of \cite{Kn}, in $\SuEu$ existence of 
trapped orbits and existence of bounded orbits are equivalent
properties.\\
In particular for $E\in \NT$ there is no rest point in $\SuEu$.
But this implies that $E$ is a regular value of $H$ on $\SuEu$,
so that $\SuEu$ is a smooth manifold.
\item
The set $\NT\subset(0,\infty)$ of non-trapping energies is open 
(see the proof of Prop.\ 2.4.1 of~\cite{DG}). 
\item
As an example of physical relevance, 
for Coulombic potentials of the form
\[V(q)= -\sum_{k=1}^n \frac{Z_k}{\|q-s_k\|}\qquad
\big(q\in\bR^3\setminus\{s_1,\ldots,s_n\}\big)
\]
in the {\em repelling} case ($Z_k<0$) there exists an interval
$(0,E_0)\subset\NT$ (see Sect.\ 5 of \cite{CJK}).\\
For $n\ge 2$, independent of the signs of the charges $Z_k$,
for $s_1,\ldots,s_n\in\bR^3$ in general position there exists an interval
$(E_{{\rm th}},\infty)$ of trapping energies,
where the dynamics of the bounded orbits is homeomorphic to
the one on the suspended flow for the shift space (\ref{def:shift})
(see \cite{Kn2}, Thm.\ 12.8).
\end{enumerate}
\end{remarks}
For $E\in \NT$ the {\em asymptotic directions}
\[\hat{p}^\pm:\SuEu\ar S^{d-1}\qmbox{,} 
\hat{p}^\pm(x):=\lim_{t\ar\pm\infty} \frac{\p(t,x)}{\sqrt{2E}}\]
and {\em impact parameters}
\[\q_\perp^{\,\pm}:\SuEu\ar\bR^d,\quad
\q_\perp^{\,\pm}(x) := \lim_{t\ar\pm\infty} 
\Big(\q(t,x)-\LA\q(t,x),\hat{p}^\pm(x)\RA \hat{p}^\pm(x)\Big)\]
are continuous $\Phi^t$-invariant functions (see \cite{DG}, Thms.\
2.5.2 and 2.7.2).

By its definition, the impact parameter
is orthogonal to the asymptotic direction, and for 
non-trapping energies $E\in \NT$ we obtain homeomorphisms
\[A_E^\pm :\SuEu/\Phi^\bR\ar N:=T^*S^{d-1}\qmbox{,} 
[x]\mapsto\big (\q_\perp^{\,\pm}(x),\hat{p}^\pm(x)\big). \]
between the space of unbounded orbits and $N$.
For $E\in\NT$ the {\em scattering map} 
\beq
S_E=(\Q_E,\hat{P}_E):=A_E^+\circ (A_E^-)^{-1}: N\ar N 
\Leq{sc:map}
is a homeomorphism
of the symplectic manifold 
$(N,\omega_N)$ and in fact a symplectomorphism, as follows from 
\cite{DG}, Thm 2.7.11.\\ 
In particular for each initial direction $\theta\in S^{d-1}$ the 
restriction 
\[\hat{P}_{E,\theta}: T^*_{\theta}S^{d-1}\ar S^{d-1}\]
of the {\em final direction map} $\hat{P}_E:=\pi_{S^{d-1}}\circ S_E:N\ar S^{d-1}$ is 
continuous. 
\begin{lemma}\label{lem:fd:map}
For all dimensions $d\geq 2$, energies $E>0$ and directions $\theta\in S^{d-1}$
\[\lim_{\|\q_\perp\|\ar\infty}\hat{P}_{E,\theta}(\q_\perp) = 
\theta.\]
\end{lemma}
{\bf Proof.}
By continuity of $(q_\perp^-,\hat{p}^-):\Sigma_E^u\to N=T^*S^{d-1}$ and
compactness of the interaction zone $\cI Z(E)$ there is an $R>0$ with the property 
that for $\|q_\perp\|>R$ the orbit $(A_E^-)^{-1}(q_\perp,\theta)\subset\Sigma_E^u$
does not intersect the compact in $\Sigma_E^u$ lying over $\cI Z(E)$.

Thus there is exactly one point 
$x(q_\perp)\equiv(p_0,q_0)\in\Sigma_E^u$ on that orbit whose 
configuration space projection $q_0$ has minimal norm. By increasing the above
$R$, that minimal distance diverges.
As $\LA p_0,q_0\RA=0$, similar to (\ref{d2qge}) we have the estimate
\[\|q(t)\|^2\geq\|q_0\|^2+\eh Et^2 \qquad(t\in\bR)\]
for the {\em whole} trajectory. Integrating the force field, that is, the 
negative acceleration, along the trajectory we get uniformly on $\Sigma_E^u$, 
by using (\ref{finite:acc})
\[\int_\bR\|F(q(t))\|\, dt\to0 \qmbox{for} \|q_0\|\to\infty.\]
Thus in this limit the change of velocity and of direction go to zero.\hfill $\Box$\\[2mm]
By one-point compactification $\Big( T^*_{\theta}S^{d-1}\cup\{\infty\} \Big) \cong S^{d-1}$ 
of that $(d-1)$-dimensional vector space
we may thus extend $\hat{P}_{E,\theta}$ uniquely to a map 
\beq
\hat{\bf P}_{E,\theta}:
 S^{d-1}\ar S^{d-1}\qquad (E\in\NT,\theta\in S^{d-1}).
\Leq{P:E:theta}
which is jointly continuous in its argument and parameters. 
The choice of an orientation on the sphere
fixes an orientation of the cotangent space $T^*_{\theta}S^{d-1}$, too,
and we denote by
\[{\rm deg}(E):= {\rm deg}(\hat{\bf P}_{E,\theta})\]
the topological degree of this map.

In general the degree of a map
$f\in C^1(S^{d-1},S^{d-1})$ is given by 
\[{\rm deg}(f)=\sum_{x\in f^{-1}(y)}\sign\det(Df(x)),\]
evaluated at an arbitrary regular value $y$ of $f$. Then 
this definition is uniquely extended to $C(S^{d-1},S^{d-1})$
(see, e.g., Hirsch \cite{Hi}, Sect.\ 5.1). 

In our case the degree is independent of the choice of orientation on $S^{d-1}$.
By joint continuity of 
$\hat{\bf P}_{E,\theta}$ in its argument and parameters it is also
independent of the choice of
initial direction $\theta$. So the 
{\em non-trapping degree}
\[{\rm deg}:\NT\ar\bZ\]
is well-defined and locally constant on the (open) set of non-trapping energies.

\begin{figure}[h] %
\centerline{
\epsfig{file=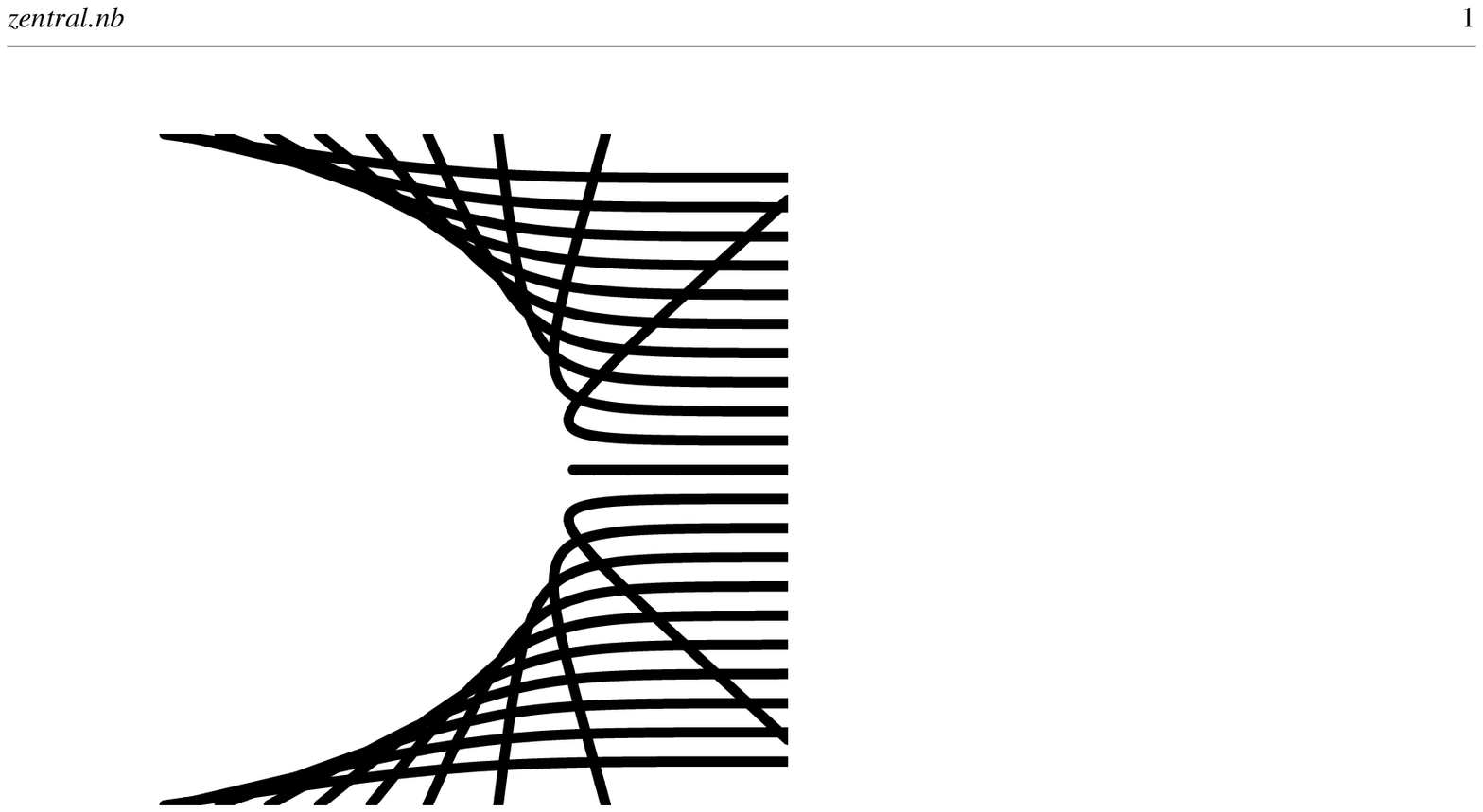,height=4cm,bbllx=90,bblly=500,bburx=320,bbury=710, 
clip=}%
\hspace*{2mm}
\epsfig{file=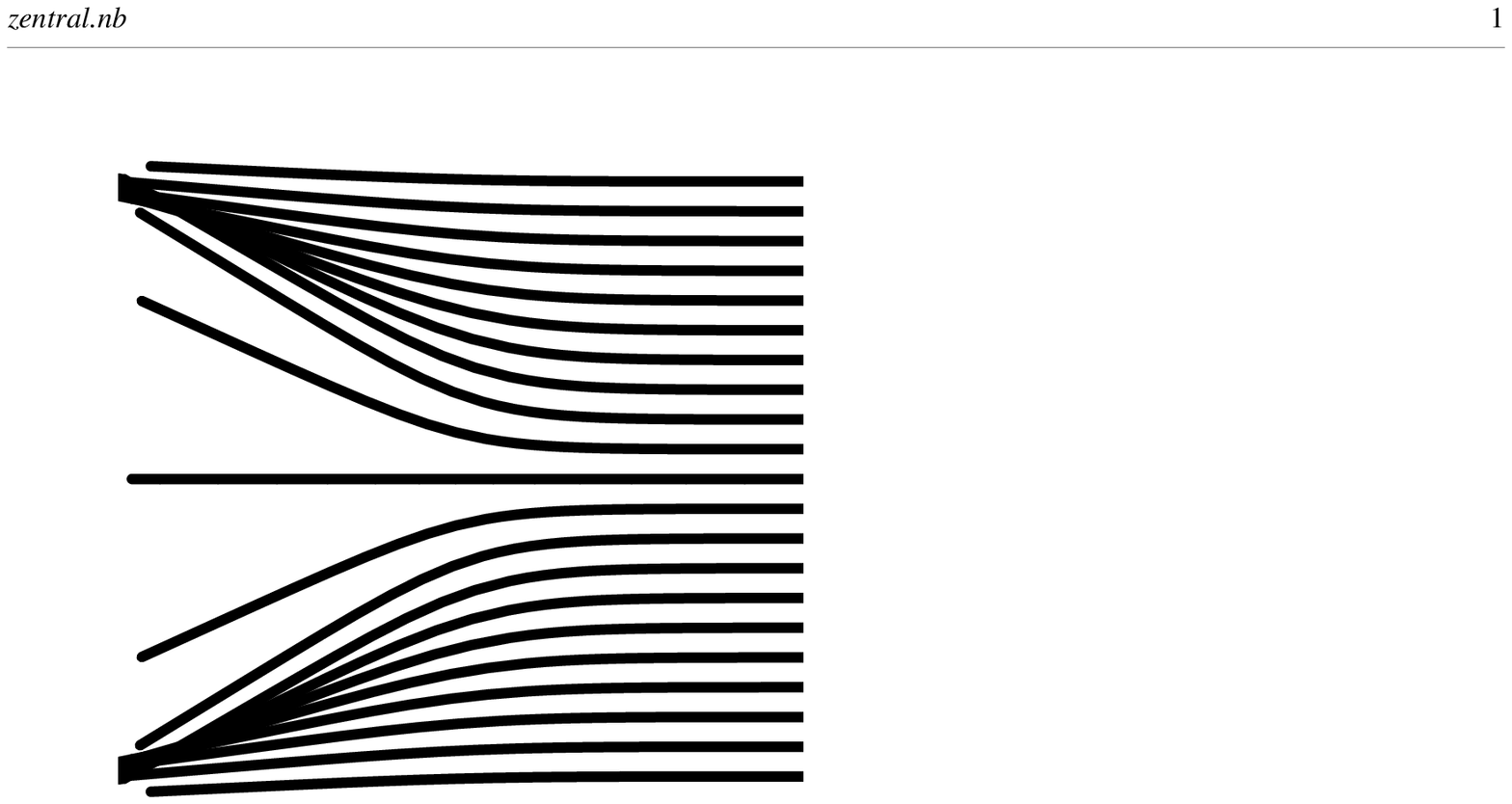,height=4cm,bbllx=90,bblly=500,bburx=320,bbury=710, 
clip=}%
\hspace*{2mm}
\epsfig{file=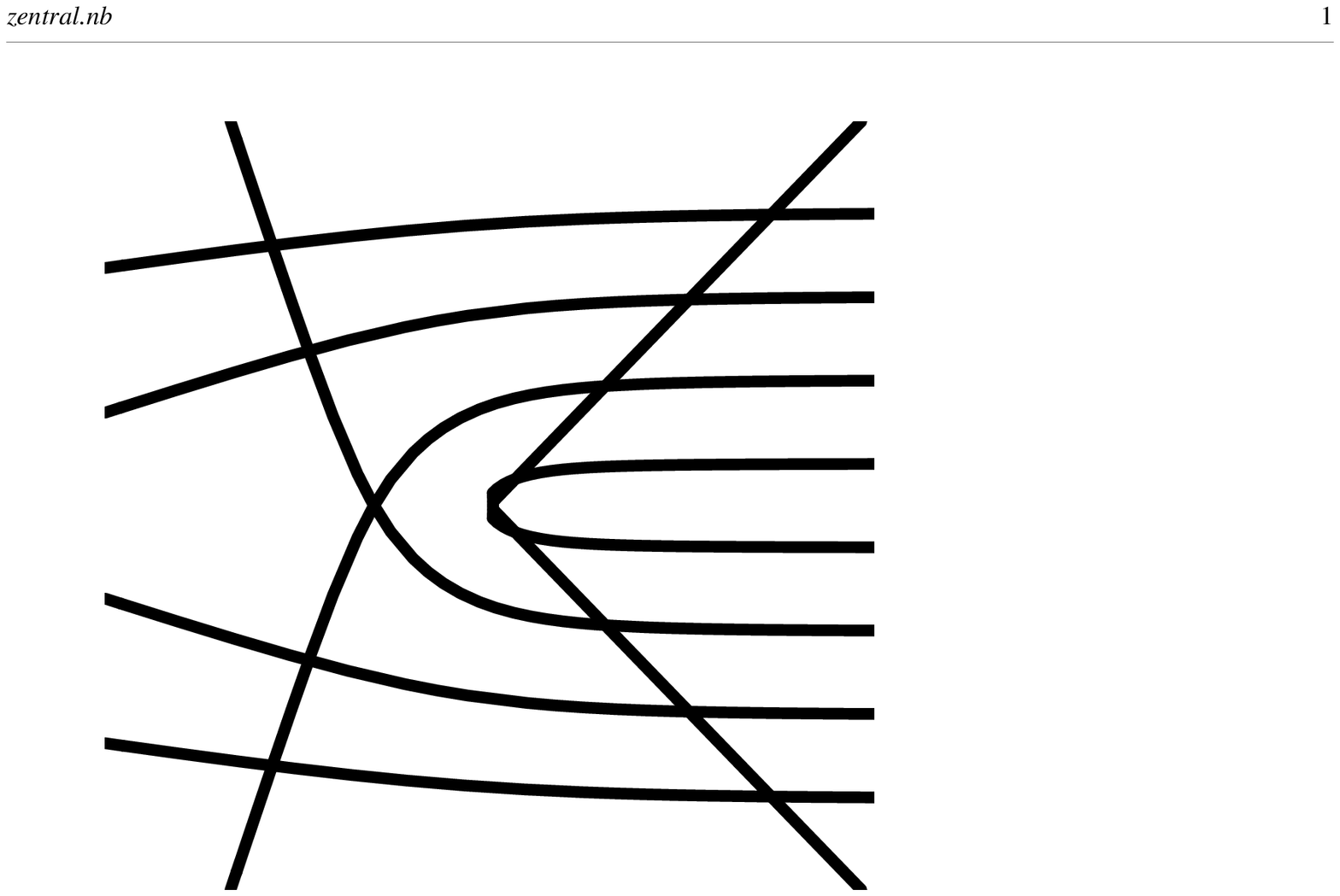,height=4cm,bbllx=90,bblly=450,bburx=360,bbury=710, 
clip=}%
}
\caption{2D scattering with degrees 1 (left), 0 (center) and -1 (right) }
\label{b1}
\end{figure}

In \cite{Kn} the degree was calculated for {\em centrally symmetric}
($V(q)=\tilde{V}(\|q\|)$) potentials, with the following results
for regular values $E$ of $V$:

\begin{itemize}
\item
For non-singular $V$
\[{\rm deg}(E) = \l\{
\begin{array}{cl}
1&,\,E\in(0,\Vmax) \\
0&,\, E\in(\Vmax,\infty)
\end{array}\ri..\]
Here
$\pa\cR_E^u$ is homeomorphic to a $(d-1)$--sphere if $E\in(0,\Vmax)$ 
and $\pa\cR_E^u=\emptyset$ for $E\in(\Vmax,\infty)$.

\item
For singular $V$ of the form $V(\q) = -Z|\q|^{-2n/(n+1)}$
all energies $E>0$ are non-trapping and 
\beq
{\rm deg}(E) = \l\{
\begin{array}{cl}
-n&,\,  d \ {\rm even}\\
\eh(1-(-1)^n)&,\, d \ {\rm odd}\end{array}\ri..
\Leq{sing:deg}
\end{itemize}
This is illustrated in Figure \ref{b1}.\\[-3mm]

In the present article we will show 
that 
\begin{itemize}
\item
also in the case of 
general non-singular potentials $V$ only the two cases 
$\pa\cR_E^u=\emptyset$ and  $\pa\cR_E^u\cong S^{d-1}$
are 
compatible with $E$ being a non-trapping energy
(Thm.\ \ref{thm:empty:sphere}), and that in this 
case $\deg(E)$ coincides with the above values (Thm.\ \ref{thm:reg}). 
\item
Likewise, it will be shown in Thm.\ \ref{thm:sing} that 
the degree formula (\ref{sing:deg}) remains true for
arbitrary smooth perturbations of the singular potential $V$ and
energies $E\in\NT$.
\end{itemize}
We set the stage by parametrizing the sets
\[L^-_{E,\theta}:=\big(\hat{p}^-\big)^{-1}(\theta)\qquad (\theta\in S^{d-1})\] 
of phase space points in $\SuEu$ with initial direction $\theta$. 
By Theorem 2.7.2 of \cite{DG}, 
given initial values $x_0,x\in L^-_{E,\theta}$, 
the limit 
\[\lim_{t\ar-\infty} \big( q(t,x)- q(t,x_0)\big)\] 
exists, 
and there exists a unique time $T_{x_0} (x)\in\bR$ such that 
\[\lim_{t\ar-\infty}\LA q\big(t+T_{x_0} (x),x\big)- q\big(t,x_0\big), \p^-(x_0)\RA =0,\]
thus asymptotically {\em synchronizing} the two trajectories.

For all $q_\perp\in T^*_{\theta}S^{d-1}$ there is a 
 unique phase space point $X=X(q_\perp,\theta)$ in
 $L^-_{E,\theta}$ with 
\[\lim_{t\ar-\infty}  \big(q(t,X)- q(t,x_0) \big) = q_\perp.\]
This gives a parametrization 
\beq
I_{\theta}:\bR\times T^*_{\theta}S^{d-1}\ar\SuEu\qmbox{,} 
(t,q_\perp) \mapsto \Phi^t\circ X(q_\perp,\theta)
\Leq{I:theta}
of the embedded $C^1$--submanifold $L^-_{E,\theta}\subset \SuEu$ by
a $d$-dimensional vector space. 
$L^-_{E,\theta}$ approximates 
for $t\ar-\infty$ the affine Lagrange space  
\[\l\{ (p,q)\in T^*M \mid p= \sqrt{2E}\ \theta\ri\}\]
in the $C^1$--sense 
\[\lim_{t\ar -\infty} p\big(t,X(q_\perp,\theta)\big)= 
\sqrt{2E}\ \theta\qmbox{,}
\lim_{t\ar -\infty}  D_{(t,q_\perp)}p\big(t,X(q_\perp,\theta)\big)=0 \] 
(see \cite{DG}, Thm. 2.7.1.).
Since $L^-_{E,\theta}$ is also
invariant under the symplectomorphisms $\Phi^t$, it is thus a Lagrange manifold, too.

\section{The Case of Regular Potentials}
%
In this section we consider potentials $V\in C^2(\bR^d,\bR)$
satisfying the long range estimate (\ref{finite:acc}),
and energies $E>0$. 
Then $V^{-1}(E)\subseteq \pa\cR_E^u$ for the unbounded component 
$\cR_E^u$ of Hill's region.

\begin{theorem} \label{thm:reg}
For non-trapping energies $E\in\cN\cT$ of $V$ 
the following holds true:
\begin{enumerate}
\item
if $\pa\cR_E^u=\emptyset$ and $d\geq2$, then $\deg(E)=0$.
\item
if $\pa\cR_E^u\cong S^{d-1}$, then $\deg(E)=1$.
\end{enumerate}
\end{theorem}
\begin{remark}
Note that only these two cases can arise for non-trapping
energies, see Thm.\  \ref{thm:empty:sphere}.
\end{remark}
{\bf Proof.}
$\bullet$
In case 1) we have $\|p|=\sqrt{2(E-V(q))}>0$ for all $(p,q)\in \SuE$ 
and thus use the continuous map (with $x^{\wedge} := x/\|x\|$)
\beq
\hat{p}:\SuE\ar S^{d-1}\qmbox{,} (p,q)\mapsto{p}^{\wedge},
\Leq{hp}
and (similar to (\ref{I:theta})) the reparametrized flow on $\SuE$
with initial direction $\theta$
\[\tilde{I}_{\theta}:T^*_{\theta}S^{d-1}\times(-1,1)\ar \SuE\qmbox{,}
(q_\perp,t)\mapsto\Phi\l(\tanh^{-1}(t),X(q_\perp,\theta)\ri).
\]
We uniquely extend their composition
\[\tilde{h}:T^*_{\theta}S^{d-1}\times(-1,1)\ar S^{d-1}\qmbox{,}
\tilde{h}= \hat{p} \circ\tilde{I}_{\theta} \]
to a map $h:S^{d-1}\times[-1,1]\ar S^{d-1}$, with
compactification 
$\Big( T^*_{\theta}S^{d-1}\cup\{\infty\} \Big) \cong S^{d-1}$,
\beq
h(\infty,t):=\theta\qmbox{,} 
h(x,-1):=\theta\qmbox{and} h(x,1):=\hat{\bf P}_{E,\theta}(x).
\Leq{hh}
By Lemma \ref{lem:fd:map} $h$ is continuous. So the restriction maps $h\rstr_{S^{d-1}\times\{i\}}$ $(i=-1,1$) are homotopic.
Thm.\ 1.6 of \cite{Hi}, Sect.\ 5.1, gives the middle equality in
\[\deg(E) = \deg\l(h\rstr_{S^{d-1}\times\{1\}}\ri)= 
\deg\l(h\rstr_{S^{d-1}\times\{-1\}}\ri) = 0,\] 
whereas the right equation follows from constancy of  $h\rstr_{S^{d-1}\times\{-1\}}$.\\
$\bullet$
In the second case ($\pa\cR_E^u\cong S^{d-1}$) the statement follows for $d=1$ trivially from the 
definition of the degree of a map $S^0\to S^0$.
So we assume $d\ge2$.\\  
Like in (\ref{hh}), we construct a homotopy 
$h$ which interpolates between $\hat{\bf P}_{E,\theta}$ and
an explicit map on the sphere whose degree we can determine.

Now (\ref{hp}) is not defined everywhere, and
we use the bounded smooth map
\[\tilde{p}:\SuE\ar \bR^{d}\qmbox{,} (p,q)\mapsto\frac{p}{\sqrt{2E}}\]
instead of $\hat{p}$. 
Note that $\|\tilde{p}(p,q)-\hat{p}(p,q)\|\ar 0$ uniformly in
$(p,q)$ as $\|q\|\ar\infty$.

Without loss of generality, we assume
that $0\in\bR^d\setminus \cR_E^u$. Then we have a smooth ($C^2$) map
\[{\Dir}:\SuEu \ar S^{d-1}\qmbox{,} (p,q)\mapsto
q^{\wedge}=\frac{q}{\|q\|}, \]
and unconditionally 
$\lim_{t\ar\infty} {\Dir} \circ \Phi(t,x)=\hat{p}^+(x)$, with 
asymptotic direction $\hat{p}^+$.\\
Unlike in case 1), we interpolate between ${\Dir}$ and $\tilde{p}$, using 
as parameter the continuous distance function 
\[{\rm Dist}_r: \SuEu\ar [0,1)\mbox{ , }
(p,q)\mapsto \tanh\l(\inf\l\{\|q-c\,\theta\|/r\mid c\le0\ri\}\ri)\qquad(r>0)\]
from the incoming axis defined by the initial direction $\theta\in
S^{d-1}$, with a suitable parameter $r$.\\
More precisely, with 
the impact parameter w.r.t.\ the initial direction $\theta$
\[\q_{\perp,\theta}:\SuEu \ar T^*_{\theta}S^{d-1}\qmbox{,}
(p,q) \mapsto q-\LA q,\theta\RA\theta,\]
$\LA x\RA:=x/\sqrt{\|x\|^2+1}$ and $t_-:=\max(-t,0)$
we define the map 
\beqno
\tilde{h}&:&T^*_{\theta}S^{d-1}\times(-1,1)\ar S^{d-1}\\
\tilde{h}(x,t)&= &
\frac{\Big( {\rm Dist}_r \cdot\tilde{p}  + (1-{\rm
Dist}_r)\cdot({\Dir}+ t_-\LA \q_{\perp,\theta}\RA)^{\wedge}\Big) \circ \tilde{I}_{\theta}(x,t)}
{\l\| \Big( {\rm Dist}_r \cdot\tilde{p}  + (1-{\rm
Dist}_r)\cdot({\Dir}+ t_-\LA \q_{\perp,\theta}\RA)^{\wedge} \Big) \circ \tilde{I}_{\theta}(x,t)
\ri\|}. 
\eeqno
$\bullet$
We begin by showing that
this is well-defined for $r$ large.

First, $\Dir+ t_-\LA q_{\perp,\theta}\RA\ne0$ since in general 
$\|\LA x\RA\|<1$ so that
$\|t_-\LA q_{\perp,\theta}\RA\|<1$.
Thus the numerator and denominator of $\tilde{h}$ are well-defined.\\
The denominator 
could only vanish on the hypersurface defined by $\Dist_r
^{-1}(1/2)$. 
Projected to configuration space, this consists of the union of\\ 
\hspace*{3mm} the hemisphere 
$\l\{q\in\bR^d\mid \|q\|=r\tanh^{-1}(\eh)\,, \,\LA q,
\theta\RA\ge0\ri\}$\\
\hspace*{3mm} and the cylinder
$\l\{q\in\bR^d\mid 
\|q- \langle q,\theta \rangle\theta\|=r\tanh^{-1}(\eh)\,, \,\langle q,\theta\rangle<0\ri\}$.

\begin{itemize}
\item[-]
For $(p,q)=\tilde{I}_{\theta}(x,t)$ projecting to the hemisphere we 
have
$\LA\theta,\hat{q}\RA \geq0$, $\LA\theta,q_{\perp,\theta}\RA =0$,
and for all $\vep>0$ $\LA\theta,
\hp\RA\geq-\vep$ if $r(\vep)$ is large. 
The last statement follows since for orbits in $\SuEu$ 
\begin{itemize}
\item
not intersecting a ball of radius $R>\Rvir$ in configuration space 
the change of maximal momentum is uniformly going to zero as
$R\ar\infty$, 
\item
whereas for the orbits intersecting in configuration space that 
interaction zone 
and then the hemisphere of radius $r>R$ at $(p,q)\in \SuEu$  
the difference 
$\|\hp-\hq\|$ uniformly goes to zero as $r\ar\infty$. 
\end{itemize}
\item[-]
On the cylinder we have 
\begin{itemize}
\item
a similar inequality for the outgoing parts of 
those orbits which have intersected the interaction zone. 
\item
For the incoming parts
\[\{ \Phi(t,x) \mid q((-\infty,t],x)\cap \IZ(E)=\emptyset\}\]
of the orbits 
\[\LA\theta,\hp\RA\geq1-\vep\qmbox{
and} \LA\theta,(\hq+q_{\perp,\theta})^{\wedge}\RA=
\frac{\LA\theta,\hq\RA}{\|\hq+q_{\perp,\theta}\|}\geq-1/2,\]
since for radius $r>1$ the denominator $\|\hq+q_{\perp,\theta}\|>2$.
\end{itemize}
\end{itemize}
So in both parts of the hypersurface the denominator of $\tilde{h}$
does not vanish.\\
$\bullet$
We now consider the limit behaviour of $\tilde{h}$ and define
an extension
\[h:S^{d-1}\times[-1,1]\ar S^{d-1}\]
of $\tilde{h}$ using these limits.
\begin{itemize}
\item[-]
$h(x,1):=\lim_{s\to1}\tilde{h}(x,s)=\hat{\bf P}_{E,\theta_-}(x)$.
Namely, in the large time limit $t_-$ vanishes, and
$\lim_{s\to1} \tilde{p}\circ \tilde{I}_{\theta}(x,s)= \lim_{s\to1} {\Dir}\circ \tilde{I}_{\theta}(x,s)$.
\item[-]
$h(\infty,s):=\lim_{\|x\|\to\infty}\tilde{h}(x,s)=\theta\qquad(s\in(0,1))$,\\
since then 
$\lim_{\|x\|\to\infty}{\rm Dist}_r \circ \tilde{I}_{\theta}(x,s)=1$.
\item[-]
The limit $h(x,-1)$ of early times is given by
\[\lim_{s\to-1}\tilde{h}(x,s)=
\l(\tanh\l(\frac{\|x\|}{r}\ri)\theta+\l(1-
\tanh\l(\frac{\|x\|}{r}\ri)\ri)(x-\theta)^{\wedge}\ri)^{\wedge}.\]
This is a continuous map $T^*_{\theta}S^{d-1}\ar
S^{d-1}$. After compactification
to a continuous map $S^{d-1}\ar
S^{d-1}$ (setting $h(\infty,-1):=\theta$) it is of degree one, as follows from linearization at 
the unique preimage $0$ of $-\theta$.
\end{itemize}
So similar to 1), by one-point compactification of the cotangent spaces 
we can uniquely extend $\tilde{h}$ to a continuous map
\[h:S^{d-1}\times[-1,1]\ar S^{d-1}\qmbox{, with} 
h\rstr_{S^{d-1}\times\{1\}}=\hat{\bf P}_{E,\theta}.\]
The two restriction maps $h\rstr_{S^{d-1}\times\{i\}}$ are homotopic 
so that 
\[\deg(E) = \deg\l(h\rstr_{S^{d-1}\times\{1\}}\ri) = \deg\l(h\rstr_{S^{d-1}\times\{-1\}}\ri) = 1.\] 
This shows the validity of the second claim.\hfill $\Box$
\begin{remarks}\label{rem:billard}
\begin{enumerate}
\item
For the class of regular potentials
meeting the inequality
\beq
\LA\q,\nabla V(\q)\RA \le 0\qquad (q\in \bR^d),
\Leq{star1}
$\NT\supseteq\{E>0\mid E\mbox{ regular value of } V\}$, as can be seen by comparison with the 
virial identity (\ref{virial}).
\item
We may also consider scattering by an obstacle $B\subset \bR^d$
diffeomorphic to a $d$--dimensional ball. 
Then $\pa B\cong S^{d-1}$
and we have the Gauss map ${\bf n}:\pa B\ar S^{d-1}$.
We use the cotangent bundle $T^*(\pa B)$ of the boundary 
of the obstacle to describe the reflection data. 
Using the euclidean metric on configuration space,
$T^*(\pa B)$ is considered as the $2(d-1)$--dimensional
submanifold of $T^*\bR^d$ annihilating ${\bf n}$. 

Without loss of generality we fix the value of the Hamiltonian
\[H:T^*M\ar \bR\qmbox{,}H(p,q)=\eh\|p\|^2\]
on $M:=\overline{\bR^d\setminus B}$ to be $\eh$ so that 
$\|p\|=1$. 
This implies that the tangential component 
$\p-\LA \p , {\bf n}(\q)\RA{\bf n}(\q)$ 
of the momentum $\p$ is contained in the unit disk
of $T^*_\q (\pa B)$. Then scattering means to invert the normal
component and to leave the tangential component invariant.

As that normal component vanishes for solutions tangential to $B$, 
the scattering map is still continuous (though not
continuously differentiable).
So if we assume that the obstacle is non-trapping, then we get index 1, by the same
argument as in the above theorem \ref{thm:reg}.
As an example, the non-trapping condition is met
if $B$ is star-shaped, since then every reflection at a point
$q\in\pa B$ increases the
value $\LA\q(t),\p(t) \RA$, in analogy to  (\ref{star1}).
\end{enumerate}
\end{remarks}
%
\section{The Case of Singular Potentials} \label{sect:sing}
%
We now treat the case of a singular potentials of the form
\beq
V(q) = -\frac{Z}{\|q-s\|^{\alpha}}+W(q) \qquad \big(\alpha\in(0,2)\big),
\Leq{V:form}
introduced in Sect.\ \ref{sect:LRP}.
Due to the singularity at the origin the Hamiltonian flow in the phase space 
$T^*M$ is incomplete. 
However, for certain values of $\alpha$ 
this flow can be completed by phase space extension. 
Then the regularization is essentially unique. 

For different regularization schemes of the representative Kepler problem
(with or without time change and change of phase space
dimension) consult Chapter II.3.4 of Cushman and Bates \cite{CB},
and Chapter 5 of Cordani \cite{Co}. 
\begin{prop} \label{deg:sing}
For $d\ge2$ the following statements are equivalent:
\begin{enumerate}
\item
$\alpha=2n/(n+1)$ for $n\in\bN$.
\item
The phase space $T^*M$ can be extended to a $2d$--dimensional
symplectic manifold 
$(P,\omega)$, with a $C^1$ flow $\Phi:\bR\times P\ar P$, extending the 
incomplete Hamiltonian flow generated by (\ref{def:H}).
\end{enumerate}
Moreover in this case $P$ is a union 
$P = T^* M\,\dot{\cup}\, (\bR\times S^{d-1})$ and can be given the structure
of a $2d$-dimensional symplectic manifold $(P,\omega)$, extending 
$(T^*M,\omega_0)$.
$H:T^*M\ar\bR$ then extends to a function in $C^2(P,\bR)$,
also denoted by $H$, having the same regular points, and its
Hamiltonian flow is $\Phi:\bR\times P\ar P$.
\end{prop}
{\bf Proof.}
$\bullet$
We denote the maximally extended (incomplete) Hamiltonian flow on 
$T^*M$ by
$\hat{\Phi}:D\ar T^*M$, with open domain $D\subset \bR\times T^*M$ on extended
phase space.
As follows from  general theory of o.d.e., $D$ is of the form
\[D =\big\{(t,x)\in \bR\times T^*M\mid t\in t\in(T^-(x),T^+(x))\big\}\]
with escape times $T^-:T^*M\to[-\infty,0):=\{-\infty\}\cup(-\infty,0)$ upper semicontinuous and 
$T^+:T^*M\to(0,\infty]$ lower semicontinuous. 

By reversibility of the flow we consider only $T^+$.
Like in Thm.\ 3.1 of \cite{MG} 
we conclude that for $T^+(x)<\infty$ we have a collision at time
$T^+(x)$, that is
\[\lim_{t\nearrow T^+(x)}q(t,x)=s.\]
Without loss of generality we assume $s=0$. Similarly we assume
that $Z=1$, using a rescaling.\\
$\bullet$
We first assume that $W=0$ in (\ref{V:form}). 
Then due to the centrally symmetric form
of $V$ every solution curve $t\mapsto q(t,x_0)$ in $M$ with initial 
conditions $x_0=(p_0,q_0)\in T^* M$ lies in the 
plane (or line) spanned by $p_0$ and $q_0$.\\
$\bullet$
So we can assume $d=2$ for the moment. The angular momentum 
\[L:T^* M\ar\bR\qmbox{,}L(p,q)=q_1p_2-q_2p_1\]
is conserved by the maximally extended Hamiltonian flow 
$\hat{\Phi}:D\ar T^*M$.
For a trajectory with energy $E$ and value $l$ of $L$
we calculate the total deflection angle $\Delta\vv(E,l)$, 
as seen from $s$.

Considering for a moment an arbitrary centrally symmetric potential
$V(\q)=\tilde{V}(\|\q\|)$
and for $l\neq 0$ its {\em effective potential} $\tilde{V}_l$ 
(with $\tilde{V}_l(r):=\tilde{V}(r)+\frac{l^2}{2r^2}$), there may or may
not be a largest $r>0$ with $\tilde{V}_l(r)=E$, then called  
the {\em pericentral radius} $\rmin$.
In this case 
we have (see Chapter 2.8 of Arnold \cite{Ar})
\beq
\Delta\vv(E,l)=2\int_\rmin^\infty \frac{\dot{\vv}}{\dot{r}}\, dr=
2\int_\rmin^\infty \frac{l/r^2}{\sqrt{2(E-\tilde{V}_l(r))}}\, dr.
\Leq{Delta:phi}
Setting $\tilde{V}(r):= -r^{-\alpha}$ with $\alpha\in(0,2)$,
we see that $\rmin$ is well-defined and non-zero for $l\neq0$. 
Substituting 
$v:=\frac{(|l|/\sqrt{2})^{1/(1-\alpha/2)}}{r}$,
we obtain
\beq
\Delta\vv(E,l)=
\sign(l)\,2\int_0^{v_{\rm max}}\frac{dv}
{\sqrt{2E|l|^{\frac{\alpha}{1-\alpha/2}}2^{\frac{1-\alpha}{2-\alpha}}+
v^\alpha-v^2}} \Leq{Dvv:n}
with $2E|l|^{\frac{\alpha}{1-\alpha/2}}2^{\frac{1-\alpha}{2-\alpha}}+
v_{\rm max}^\alpha-v_{\rm max}^2=0$. Since $\alpha<2$, 
in the collision limit $l\ar 0$
the first term in the square root vanishes, and
\beq
\Delta\vv^\pm := \lim_{\pm l\searrow0} \Delta\vv(E,l)
= \pm2 \int_0^1 \frac{dv}{\sqrt{v^\alpha-v^2}} 
= \pm\frac{2\pi}{2-\alpha},
\Leq{Dphi:l}
which equals $\pm (n+1)\pi$ if $\alpha=2n/(n+1)$.\\ 
So precisely for those exponents $\alpha\in(0,2)$
that appear in our first assertion 
we have $\Delta\vv^+=\Delta\vv^-\ ({\rm mod }\, 2\pi)$.
This shows the implication 2) $\Longrightarrow$ 1).\\
Moreover for $\alpha=2n/(n+1)$ formula (\ref{Dvv:n}) equals
\[\Delta\vv(E,l)=\sign(l)\,2\int_0^{v_{\rm max}} \frac{dv}
{\sqrt{2E l^{2n}2^{(1-n)/2}+
v^{2n/(n+1)}-v^2}}\] 
which with (\ref{Dphi:l}) extends to a $S^1$-valued function of
$l$ and $E$ which is smooth even at $l=0$.
\\
$\bullet$
In order to prove the implication 1) $\Longrightarrow$ 2), we now
assume $\alpha=2n/(n+1)$ with $n\in\bN$.
Then we
can continuously regularize the collision orbits with $l=0$
after collision at time $t_0$, by setting 
\[\Big(p(t_0+\Delta t),q(t_0+\Delta t)\Big):= 
\Big((-1)^n p(t_0-\Delta t),(-1)^{n+1} q(t_0-\Delta t)\Big)\qquad(\Delta t>0).\]
Still that trajectory is undefined for time $t_0$, since
$q(t_0)\not\in M$.\\
$\bullet$
$P$ can be made a $2d$-dimensional manifold and $\Phi$
a smooth Hamiltonian flow, by using adapted coordinates
in a suitable phase space neighbourhood 
\[\hat{U}^\vep := 
\l\{\pq\in T^{*}M \l| \|\q-s\|<\vep,\,\,
\|\p\|^{2}>\frac{c_\alpha\,Z }{\|q-s\|^\alpha} \ri.\ri\}\]
with $c_\alpha:=\frac{2+\alpha}{2}\in(\alpha,2)$. 
For small $\vep$ within $\hat{U}^\vep$
\beqn
\frac{d}{dt}\LA\q-s,\p\RA &=& 
\|p\|^2-\frac{\alpha Z}{\|q-s\|^\alpha}
-\LA\q-s,\nabla W(\q)\RA\NN\\
&>&
\frac{c_\alpha-\alpha}{2}\frac{Z}{\|\q-s\|^\alpha}-\LA\q-s,\nabla W(\q)\RA>0.
\label{elapse}
\eeqn
So within $\hat{U}^\vep$ the flow is transversal to
the {\em pericentric hypersurface} 
\beq
S_0:=\{(p,q)\in T^*M\mid \LA\q-s,\p\RA =0\}.
\Leq{S0}
As $\frac{d^2}{dt^2} \|q-s\|^2=2\frac{d}{dt}\LA\q-s,\p\RA $, this inequality also shows that the point of the orbit on $S_0$ is
indeed pericentric.
Every collision orbit enters $\hat{U}^\vep$, since
\[\|p\|^2-\frac{c_\alpha Z}{\|q-s\|^\alpha}=2
\frac{(2-c_\alpha)Z}{\|\q-s\|^\alpha}+2(E-W(\q))\ar\infty\]
as $\q$ approaches $s$.\\
$\bullet$
In the present case $W=0$ we use
the following coordinates on $\hat{U}^\vep$.
\begin{itemize}
\item[-]
$H\rstr_{\hat{U}^\vep}\in C^\infty(\hat{U}^\vep,\bR)$.
The value $E:=H(x)$ of the Hamiltonian function at
$x$ is conserved by the flow. 
\item[-]
The time $T:\hat{U}^\vep\ar \bR$ 
needed to arrive at the pericentre respectively at $s$. 
As the Hamiltonian function is smooth, we have 
$\hat{\Phi}\in C^\infty(D,T^*M)$.
Furthermore by (\ref{elapse}) the flow $\hat{\Phi}$
is transversal to the smooth
pericentric hypersurface $S_0$, so that
$T$ is smooth for all points $x:=(p,q)\in\hat{U}^\vep$ 
on non-collision orbits.

Moreover, $T$ is explicitly given by the integral over inverse radial
velocity:
\beqno 
T(x)&=&\sign\l(\langle p,q\rangle\ri)
\int_{\rmin}^{\|q\|}\frac{1}{\sqrt{2\big(E-\tilde{V}_{l}(r)\big)}}\, dr,
\\ &=&\sign\l(\langle
p,q\rangle\ri)2^{-n}\int_{\frac{l^{n+1}}{2^{(n+1)/2} \|q\|}}^{v_{\rm max}}
\frac{l^{2(n+1)}dv}
{v^2\sqrt{2E l^{2n}2^{(1-n)/2}+
v^{2n/(n+1)}-v^2}}
\eeqno  
with $E:=H(x)$, $l:=L(x)$, and
similar to $\Delta\vv^\pm$, $T \in C^\infty(\hat{U}^\vep,\bR)$.
\item[-]
As in \cite{MG}, we now 
discern the cases of even resp.\ odd $n\in\bN$.\\
In each case we define a map $F \in C(\hat{U}^\vep,S^{d-1})$
in a way so that for initial conditions $x_0=(p_0,q_0)\in\hat{U}^\vep$
the direction $F(x_0)$ lies
in the two-plane (or line) spanned by $p_0$ and $q_0-s$.
\\[2mm]
For {\bf odd} $n$ and 
non-zero angular momentum of $x_0$ 
we define $F(x_0)\in S^{d-1}$
as the direction  of the pericentre of the orbit through $x_0$. For
$t_0 := T(x_0)$ 
(that is, $\Phi^{t_0}(x_0)\in S_0$)
this equals
\[F(x_0)=\frac{q({t_0},x_0)-s}{\|q({t_0},x_0)-s\|}.\]
Dividing the expression (\ref{Dphi:l}) for the limiting deflection 
angle of collision orbits by 2,
we note that for zero angular momentum of $x_0$ 
and for $\alpha=2n/(n+1)$, $n$ odd we get 
$F(x_0)=(-1)^{(n+1)/2}\frac{q_0-s}{\|q_0-s\|}$.\\
%
For {\bf even} $n$ and 
non-zero angular momentum of $x_0$ we define $F(x_0)$
as the normalized velocity 
$F(x_0):=\frac{p({t_0},x_0)}{\|q({t_0},x_0)\|}$
at the pericentre of the orbit through $x_0$.
By (\ref{S0}) this is perpendicular to the vector $q({t_0},x_0)-s$.\\
So using formula (\ref{Dphi:l})
for the limiting deflection angle of collision orbits
we note that for zero angular momentum of $x_0$ 
and for $\alpha=2n/(n+1)$, $n$ even we get 
$F(x_0)=(-1)^{n/2}\frac{p_0}{\|p_0\|}$.
\item[-]
The conserved (non-zero) value $l$ 
of the 'angular momentum vector at the pericentre' 
$L(x):=\|q({t_0},x)-s\|\, p({t_0},x)$.
$(l,\vv)$ is a point in the symplectic manifold
$T^*S^{d-1}$.
%

\end{itemize}
The collision orbits correspond to the points with $l=0$, but
$T(p,q)\neq 0$. The cylinder $\bR\times S^{d-1}$ in (\ref{def:P})
is then identified with the set of missing phase space points,
characterized by  $(l,t)=0$.\\
In the above coordinates the
flow is affine in the variable $T$ ($T\circ \Phi^t=T+t)$), 
the other variables being constants of motion.
So $\Phi^t$ can be uniquely extended to the cylinder, and the 
resulting flow on $P$ is smooth and complete.\\

That collision orbit can thus be parametrized by its energy $E\in\bR$ and,
say, initial direction $\theta\in S^{d-1}$. So by setting
\beq
P:= T^*M\,\dot{\cup}\, (\bR\times S^{d-1}),
\Leq{def:P}
we may thus regularize the motion on this new phase space and obtain a
complete, smooth  flow extending $\hat{\Phi}$
\[\Phi:\bR\times P\ar P.\]
$\bullet$
If the smooth potential $W$ in (\ref{V:form})
is non-zero, the above quantities 
$H,L$ and $F$ are not conserved. However, 
they can be used to define conserved quantities, namely their
values at the unique pericentre of the near-collision orbit. 
 
See \cite{KK}, Prop.\ 2.3 and \cite{Kn2} Thm.\  5.1
for details of the (somewhat technical) 
construction in the representative case $n=1$ of the Kepler
potential.\\
$\bullet$
We now extend the natural symplectic form on $T^*M$ to
$P$, defining it by $(\Phi^t)^*\omega_0$ on  $P\setminus T^*M$.
More precisely, by (\ref{elapse}) for any $x\in P\setminus T^*M$ there is
an open neighbourhood $U\subset P$ of $x$ 
and $t>0$ such that $V:=\Phi(t,U)\subset T^*M$. The restriction
$\Psi:=\Phi^t\rstr_V$ is 
$\omega_0$-symplectic on $V':=\Psi^{-1}(U\cap T^*M)$. We uniquely
extend $\omega_0$
to $P$ by setting $\omega\rstr_U:=\Psi^*\omega_0$
(Concrete expressions of $\omega$ in terms of local coordinates
can be found in \cite{KK}, Prop.\ 2.3 and \cite{Kn2} Thm.\  5.1).\\
$\bullet$
That the Hamiltonian function extends to a function $H\in C^2(P,\bR)$ 
having no singular points on $P\setminus T^*M$ and generating $\Phi$,
is immediate from the foregoing construction, since $H$ is one of the
coordinates used in the definition of $P$. 
\hfill$\Box$\\[2mm]

In the regularizable case for $E>V_{\max}$ the energy surface $\Sigma_E$ is a 
$(d-1)$-sphere bundle
\beq
\pi_E:\Sigma_E\to\bR^d
\Leq{p:E}
over configuration space. 
As the base $\bR^d$ is contractible, this bundle is trivial. 
The same statement applies for all directions $\theta$ 
to the induced bundles
\[\xi_\theta:=(\pi_E\circ I_{\theta})^*\pi_E:
\tilde{\Sigma}_E\to \bR\times T^*_{\theta}S^{d-1}\qquad\qquad(\theta\in
S^{d-1})\]
over the (parametrized) Lagrange manifolds $L^-_{E,\theta}$, see (\ref{I:theta}).
By definition of induced bundles (compare with \cite{Hi}, Sect.\ 4.2) the total space of $\xi_\theta$ equals
\[\tilde{\Sigma}_E:=I_{\theta}^*\Sigma_E :=
\{(x,y)\in  \bR\times T^*_{\theta}S^{d-1}\times\Sigma_E\mid
I_{\theta}(x) = \pi_E(y) \}.\]
However, if we consider the {\em local degree $e(S)$
of the section} 
\[S:M \ar \SuE,\qmbox{,} q\mapsto \l(\sqrt{2(E-V(q))}\,\,\theta,q\ri) \]
see Bott and Tu \cite{BT}, \S 11, then this is non-trivial.
By definition this is the degree of the composed map
\beq
 S^{d-1}\cong \pa B_r\stackrel{S}{\ar} \SuE\rstr_{B_r}\cong
B_r\times S^{d-1} \stackrel{\rho}{\ar} S^{d-1},
\Leq{euler}
for a ball $B_r :=\{q\in\bR^d\mid \|q-s\|\le r\}$ of arbitrary radius
$r>0$.

\begin{remark}
In the case of sphere bundles over a compact base manifold $M$, the
{\em Euler number} of the bundle is the sum of local degrees at finitely
many base points, see \cite{BT},~Thm.\ 11.16.

In this context the Euler number vanishes for $d=\dim(M)$ odd, if the
oriented
$(d-1)$--sphere bundle is the restriction of a vector bundle of rank
$d$ over $M$. 

As Formula (\ref{loc:deg}) below indicates, for our
bundle $\pi_E:\Sigma_E\to\bR^d$ this is not
the case if $n$ is odd.
\end{remark}
\begin{theorem} \label{thm:sing}
For a non-trapping energy $E\in\cN\cT$ of $V$ the following
holds true.
If $V$ is of the form (\ref{V:form}) with $\alpha=2n/(n+1)$ for $n\in\bN$, then for 
$d\geq2$
\beq
\deg(E)=e(S) = \l\{ \begin{array}{ccc} -n&,&d\ \mbox{even}\\
\frac{1-(-1)^n}{2}&,&d\ \mbox{odd}.\end{array}\ri.
\Leq{loc:deg}
\end{theorem}
{\bf Proof.}
$\bullet$
We start by calculating the local degree $e(S)$.
Without loss of generality we assume that the singularity is located
at $s=0$, with $\alpha=2n/(n+1)$.\\
$\bullet$
We consider first the case of a potential
(\ref{sing:V}) with $W=0$.
Then, as there are no (semi)-bounded orbits of positive energy, 
$\NT=\bR^+$, and we can use the formula 
\beq
\Delta\vv=\lim_{l\ar0} \Delta\vv(E,l) = n\pi
\Leq{defl:ang} 
for the limit of the total deflection angle, 
derived in (\ref{Dphi:l}). 

We define a section 
of the bundle $\pi_E:\Sigma_E\to\bR^d$ by  
\beq
T:\bR^d
\ar\Sigma_E\qmbox{,}T(q):=\l\{ \begin{array}{ccc}
\l(\sqrt{2(E-V(q))}\,\,F(\theta,q),q \ri)&, q\neq s\\
\big(E, F(\theta,q)\big)&,q=s
\end{array}\ri.,
\Leq{T}
where, similar to the proof of Prop.\ \ref{deg:sing}, $F:\Sigma_E\ar S^{d-1}$ maps $x$
to the unique pericentral direction of the orbit through $x$.
This section is continuous, the
apparent discontinuity of (\ref{T}) 
at $q=s$ being owed to the use of the local
cylinder coordinates $(E,\vv)\in\bR\times S^{d-1}$.\\
$\bullet$
Evaluating the degree of (\ref{euler}) in the limit $r\ar\infty$,
we can use (\ref{defl:ang}) to obtain the second equality in
(\ref{loc:deg}).

In the case of $d=2$ dimensions the outgoing angle  
\[\hat{{\bf P}}_{E,\theta}(q_\perp)=\theta -\Delta\vv(E,\sqrt{2E}q_\perp) 
\qquad(q_\perp\geq 0)\]
is continuous decreasing in $q_\perp$. So in this case
it follows from (\ref{Dphi:l}) that 
\beq
\int_{-\infty}^\infty \frac{d}{dq_\perp}
\hat{{\bf P}}_{E,\theta}(q_\perp)\, dq_\perp = -2\Delta\vv=-2\pi n.
\Leq{twice}
This is twice the change in direction from $\theta$ to $F(\theta,q)$,
since by symmetry the change in direction before and after the
time of pericentre are equal.
The section $T:\bR^d\ar\Sigma_E$ based on $F$
trivializes the circle bundle, and on $\pa B_r$ the difference between
the sections $S$ on $T$ is given by $q\mapsto F(\theta,q)- \theta$.

On the other hand half of (\ref{twice}) is the contribution 
of the part $\{q\in\pa B_r\mid \LA q,\theta\RA \le0 \}$ (on the left hand
side of (\ref{euler})) to $e(S)$, since it corresponds to the incoming 
parts of the orbits. By symmetry the outgoing 
parts of the orbits, corresponding to $\{q\in\pa B_r\mid \LA
q,\theta\RA \ge0 \}$,
give the same contribution.

Together this proves
\[{\rm deg}(E)= -n\qquad (E>0).\]
$\bullet$
For $d>2$ we consider a family of trajectories with fixed $E$ and $\theta$, 
whose impact parameter $\q_\perp$ varies on a one-dimensional subspace
$L\subset T^*_{\theta}S^{d-1}$. 

$\theta$ and this subspace span a 2--plane in $\bR^d$, and
$\theta^+$ lies in that plane. To avoid degeneracies we choose a
$\theta^+$ which is linear independent from $\theta$. Then there are
exactly $n$ impact parameters $\q^1_\perp,\ldots,\q^n_\perp\in L$
with $\hat{{\bf P}}_{E,\theta}(\q^i_\perp)=\theta^+$.

$[n/2]$ of them have a scalar product $\LA \q^i_\perp, \theta^+ \RA>0$,
and $\LA \q^i_\perp, \theta^+ \RA<0$ for the rest.
For the first group the restriction of the linearization of the final angle map
to the subspace $\{{v}\in T^*_{\theta}S^{d-1}\mid {v}\perp L\}$
gives a positive sub-determinant, whereas for the second group
the sign equals $(-1)^{d-2}$. So
\[{\rm deg}(E)=-\l([n/2]+(-1)^{d-2}(n-[n/2])\ri), \]
proving the second equality in
(\ref{loc:deg}).
For the case $W=0$ this also proves the first equality in
(\ref{loc:deg}), using (\ref{sing:deg}).\\
$\bullet$
Now we turn to the case of non-vanishing $W$.
The local degree $e(S)$ is independent of the radius $r>0$ in
(\ref{euler}). Evaluating $e(S)$ in the limit $r\ar0$, we see
that by smoothness of $W$ it coincides with $e(S)$, 
calculated above for the case $W=0$. This proves 
the second equality in (\ref{loc:deg}) for arbitrary $W$.

The map $(\pi_E,\hat{p}^+):\SuEu\ar \bR^d\times S^{d-1}$
is another trivialization of the bundle, and $\hat{p}^+$, 
evaluated over
the sphere $\pa B_r\times\{\theta\}$, has the degree 
$\deg(E)$. This shows in the general case that $e(S)=\deg(E)$.
\hfill $\Box$
%
\section{Projection of the Lagrange Manifold} \label{sect:lagrange}

This section applies to regular as well as to singular potentials. 

As noted in Sect.\ \ref{sect:LRP},
for $E\in\cN\cT$ and $\theta\in S^{d-1}$, the image
$L^-_{E,\theta}$ of the embedding
\[I_{\theta}:\bR\times T^*_{\theta}S^{d-1}\ar\SuEu\subset P\]
is a Lagrange manifold in phase space $P$.
We now consider the projection $\pi:P\to\bR^d$ and the composition map
\[\Pi_E:=\pi\circ I_{\theta}:\bR\times T^*_{\theta}S^{d-1}\to \cR_E^u,\]
mapping this Lagrange manifold to configuration space. This is a $C^1$-map
between $d$-dimensional $\pa$-manifolds. Moreover, it is proper, that is, 
compacts have compact preimages.

We orient the vector space $\bR\times T^*_{\theta}S^{d-1}$ 
so that for all $x\in\bR^{d-1}$
\[\det(D\Pi_E(t,x))>0 \qmbox{for} t\ll 0\]
(then, in fact, $\lim_{t\to-\infty} \det(D\Pi_E(x,t))=\sqrt{2E}$).

Then for every regular value $q$ of $\Pi_E$, we set 
\[\deg_q(\Pi_E):=\sum_{y\in\Pi_E^{-1}(q)}\sign(\det(D\Pi_E(y))).\]
This is well-defined. By properness of $\Pi_E$ and connectedness of 
$\cR_E^u$, the value does not depend on $q$. Thus we obtain an integer
\[\deg(\Pi_E)\in\bZ.\]
\begin{theorem} \label{thm:deg:2}
$\deg(E)=1-\deg(\Pi_E)$.
\end{theorem}
{\bf Proof.}
We evaluate $\deg_q(\Pi_E)$ for $q=\Pi_E(x,t)$ in the limit $t\to-\infty$.
Then, as remarked above, the point $(x,t)\in\Pi_E^{-1}(q)$ contributes a 
one to $\deg(\Pi_E)$. The other contributions come from points 
$(x',s)$ with
$s\to\infty$ as $t\to-\infty$ and $\hp^+(X(x',s))\to-\theta$.
So they correspond to the terms contributing to
\[{\rm deg}(\hat{\bf P}_{E,\theta})=
\sum_{q_\perp\in \hat{\bf P}_{E,\theta}^{-1}(-\theta)}\sign\det(D
\hat{\bf P}_{E,\theta}(q_\perp)),\]
but the signs in the sum for $\deg(\Pi_E)$ 
are reversed, since the initial
direction $\theta$ is reversed for these orbits.
\hfill $\Box$
%
\section{A Topological Criterion for Trapping} \label{sect:cobordism}
%
This section applies to regular potentials. 
The following topological criterion for trapping generalizes a
low-dimensional ($d\leq 3$) result of \cite{Kn} to arbitrary 
dimensions $d$.

\begin{theorem} \label{thm:empty:sphere}
If $E\in\NT$ for a long-range potential $V:\bR^d\ar\bR$, $d\in\bN$, then 
the boundary $\pa\cR_E^u$ of Hill's region
is empty or homeomorphic to $S^{d-1}$.
\end{theorem}
{\bf Proof.}
$\bullet$
If $E\in\NT$, then
the relative homotopy groups of Hill's region w.r.t.\ its boundary
are trivial,
\beq
\pi_k(\cR_E^u,\pa\cR_E^u) = \{e\}\qquad(k\in\{0,\ldots,d\}).
\Leq{rel:trivial}
This was shown in Thm.\ 3.2 of \cite{Kn} for short range 
smooth potentials, but the argument only involved the
dynamics within the interaction zone $\IZ(E)$ and thus generalizes to our
class of long range potentials. 

Our aim is to invoke the
$h$-cobordism theorem in order to show that (\ref{rel:trivial}) 
implies $\pa\cR_E^u\cong\emptyset$ or $\pa\cR_E^u\cong S^{d-1}$,
$\cong$ denoting existence of a homeomorphism.\\
$\bullet$
We assume that $\pa\cR_E^u\neq\emptyset$ and have to show 
$\pa\cR_E^u\cong S^{d-1}$.
First of all, for $E\in\NT$, $\pa\cR_E^u$ is a closed 
$(d-1)$--dimensional manifold, embedded in $\bR^d$:
\begin{enumerate}
\item
$E$ is a regular value of $V\rstr_{\cR_E^u}$ for otherwise
there would exist a equilibrium point of the flow, restricted to
$\SuEu$. By Remark \ref{rem:tra}.4 
this would contradict the assumption $E\in\NT$.
As $\pa\cR_E^u$ is a component of $V^{-1}(E)$, it is a
$(d-1)$--submanifold,
and being itself a boundary, without boundary.
\item
Since by assumption $\lim_{\|q\|\ar\infty} V(q)=0$ but
$E>0$ and $V\rstr_{\pa\cR_E^u}=E$, the boundary $\pa\cR_E^u$ of
Hill's region is compact, and thus closed as a manifold.
We choose $R>0$ so that it is contained in the interior
of the ball $B^d_R(0)$.
\end{enumerate}
Moreover, we can assume $\pa\cR_E^u$ to be connected, for otherwise by
curve shortening one could find a solution moving between 
two components of $\pa\cR_E^u$ (a so-called {\em brake orbit}, 
see Seifert \cite{Se} and Gluck and Ziller \cite{GZ}). 
Again by Remark \ref{rem:tra}.4, such a bounded orbit in $\SuEu$ would be an obstruction 
to our assumption $E\in \NT$. \\
$\bullet$
We want to apply the $h$-cobordism theorem (see \cite{Mi}, Thm.\ 9.1) 
to the  triad $({\cal W}; {\cal V},\pa\cR_E^u)$, with 
\[{\cal W}:= \cR_E^u\cap B^d_R(0)\qmbox{and }
{\cal V}:=\pa B^d_R(0).\] 
So the boundary of the manifold ${\cal W}$
equals $\pa {\cal W}={\cal V}\,\dot{\cup}\,\pa\cR_E^u$. 

In fact the subset ${\cal W}$ of 
$\cR_E^u$ is homotopy equivalent to  $\cR_E^u$, as
follows from the deformation retraction
\[H:[0,1]\times \cR_E^u \rightarrow \cR_E^u\mbox{ , }
H(t,x):=
\idty_{\{\|x\|<R\}}(x)\cdot x+
\idty_{\{\|x\|\geq R\}}(x)\cdot\Big((1-t)x+t\frac{Rx}{\|x\|}\Big)
\] 
between the identity 
map $\idty_{\cR_E^u}$ and the map 
\[f:\cR_E^u\rightarrow {\cal W} \qmbox{,}
x\mapsto\idty_{\{\|x\|<R\}}(x)\cdot x+
\idty_{\{\|x\|\geq R\}}(x)\frac{R\,x}{\|x\|},\] 
composed with the inclusion  ${\cal W}\rightarrow \cR_E^u$.

We can assume $d\ge4$,
since the cases $d\le 3$ have been analyzed in \cite{Kn},
using low-dimensional methods.

There are several assumptions to be checked in order to
apply that theorem.\\
$\bullet$
First we have to ascertain that 
${\cal W}$, ${\cal V}$ and $\pa\cR_E^u$ are simply connected.
This is trivial for the $(d-1)$-sphere ${\cal V}$. 

We claim that $\pi_{1} ({\cal W} , {\cal V}) = \{e\}$ which then will imply
$\pi_{1} ({\cal W}) = \{e\}$, using the exact sequence 
\beq
\{e\}=\pi_{1}({\cal V})\ar
\pi_{1}({\cal W})\ar\pi_{1}({\cal W},{\cal V}).
\Leq{first:exact}
A second implication will be that also $\pa\cR_E^u$, the boundary of
Hill's region, is simply connected. 
Here one takes the exact sequence 
\beq \{e\}=\pi_2({\cR_E^u},\pa{\cR_E^u})=\pi_2({\cal W},\pa{\cR_E^u})\ar 
\pi_{1}(\pa\cR_E^u)\ar \pi_{1}({\cal W})=\{e\},
\Leq{second:exact}
the first identity being (\ref{rel:trivial}), the second using the deformation retraction $H$.\\
$\bullet$
To show vanishing of  $\pi_{1} ({\cal W} , {\cal V})$ in
(\ref{first:exact}), we construct 
for any two representatives $h_i:[-1,1]\ar {\cal W}$, $h_i(\pm1)\in {\cal V}$
of relative homotopy classes in $\pi_{1} ({\cal W}, {\cal V})$ a
homotopy 
\beq
H:[0,1]\times [-1,1]\ar {\cal W} \qmbox{with}
H\rstr_{\{i\}\times [-1,1]}=h_i\quad(i=0,1).
\Leq{Hhh}
This would be simple if we could assume that the representatives $h_i$
are solution curves of the Hamiltonian equations with energy $E$ 
(with suitable reparametri\-zations, mapping the time intervals onto
$[-1,1]$).
Then we could vary initial conditions from the one for $h_0$
to the one for $h_1$, and obtain a homotopy $H$.

What we can do instead is the following. Without loss of generality we
assume that the $h_i$ are smooth regular curves not meeting 
$\pa\cR_E^u$. 

For the scaling constants
$c_i(t):=\sqrt{2(E-V(h_i(t))}/\|\dot{h}_i(t)\|$
the initial conditions 
$x_i(t):=\big(c_i(t)\dot{h}_i(t),h_i(t)\big)$ are in $\SuEu$.
The maps 
\[\tilde{g}_i:D_i\ar \SuEu\qmbox{,}
\tilde{g}_i(t,s):=\Phi^s(x_i(t))\] 
with domains
$D_i:=\{(t,s)\in [-1,1]\times \bR\mid s\in [T^-_i(t),T^+_i(t)]\}$
are $C^1$-smooth, and the functions $T^-_i\leq0\leq T^+_i$ are uniquely defined by
the conditions $\tilde{g}_i\big(t,T^\pm_i(t)\big)\in {\cal P}_E^\pm$
for the disjoint hypersurfaces
\[{\cal P}_E^\pm:=
\big\{(p,q)\in \SuEu\mid \|q\|=R, \pm \LA p,q\RA > 0\big\}\] 
over ${\cal V}$.
Namely, the flow line through $x_i(t)\in\SuEu$ intersects both surfaces, since 
$E\in\NT$. 
Furthermore by the virial inequality (\ref{qp}) both are intersected at most once 
by a flow line. By the same argument we get transversality of the
intersection and thus $C^1$-smoothness of $T^\pm_i$.

As ${\cal P}_E^-$ is connected, there is a homotopy
$\psi:[0,1]\times [-1,1]\ar {\cal P}_E^-$ with
\[\psi(i,t)=\Phi\big(T^-_i(t),x_i(t)\big)\qquad
\big(i\in\{0,1\}, t\in[-,1,1]\big).\]
This gives rise to a continuous map 
$\rho:D_\rho\ar\SuEu,\ \rho(z,t,s):=\Phi^s(\psi(z,t))$, on
\[D_\rho := \big\{(z,t,s)\in [0,1]\times [-1,1]\times [0,\infty)\mid 
s\le S(z,t)\big\},\]
with the continuous Poincar\'{e} time $S$
uniquely given by $\Phi\big(S(z,t),\psi(z,t)\big)\in {\cal P}_E^+$.

For a homeomorphism 
\[K:[0,1]\times [-1,1]\times [-1,1]\to D_\rho
\qmbox{,}(z,t,s)\mapsto \big(z,t,k(z,t,s)\big)\]
with $k(z,t,-1)=0$, $k(z,t,+1)=S(z,t)$ and
$k(i,t,t)=-T^-(i,t)$
the configuration space projection (see (\ref{p:E}))
\[F:=\pi_E\circ\rho\circ K:[0,1]\times [-1,1]\times [-1,1]\to{\cal
W}\]
has the property 
\[H(i,t)=h_i(t)\qquad \big(i\in\{0,1\},t\in[-1,1]\big)\] 
for $H(z,t):=F(z,t,t)$. So $H$ is the desired homotopy from 
(\ref{Hhh}).\\
%
%
%
%
%
$\bullet$
As a second condition of the $h$-cobordism theorem we have to ascertain that the relative homology
\[H_*({\cal W},\pa\cR_E^u) = 0.\]
This follows from $H_*(\cR_E^u,\pa\cR_E^u)=0$ since 
${\cal W}$ was shown to be homotopy equivalent to  $\cR_E^u$.
By the relative Hurewicz isomorphism theorem
(\cite{Sp}, Chapter 7.5) vanishing of that
relative homology follows from (\ref{rel:trivial}).\\
$\bullet$
The last condition of the $h$-cobordism theorem
is that $\dim({\cal W})\ge 6$. In fact a version of the $h$-cobordism theorem
in the homeomorphic category works for $d=\dim({\cal W})\ge 5$, see 
Sect.\ 7.1 in the book \cite{FQ} by {Freedman} and {Quinn}.\\
In these cases its conclusion is that the following triads are
homeomorphic:
\[({\cal W}; {\cal V},\pa\cR_E^u) \cong 
\Big({\cal V}\times[0,1], {\cal V}\times\{0\},{\cal V}\times\{1\}\Big);\]
in particular $\pa\cR_E^u$ is homeomorphic to a $(d-1)$--sphere.
\\
$\bullet$
The case not covered by the cobordism theorem is the one of dimension $d=4$.\\
Here the celebrated proof of the Poincar\'{e} conjecture
by Grigori Perelman allows us to 
conclude that $\pa\cR_E^u$ is homeomorphic to $S^3$, since
by (\ref{second:exact}) $\pa\cR_E^u$ is a connected closed 3-manifold
with fundamental group $\pi_1(\pa\cR_E^u)=\{e\}$.\hfill $\Box$
%

\end{document}